\documentclass[conference]{IEEEtran}
\IEEEoverridecommandlockouts

\usepackage[ruled,vlined]{algorithm2e}
\usepackage{cite}
\usepackage{amsmath,amssymb,amsfonts}
\usepackage{algorithmicx}
\usepackage{graphicx}
\usepackage{textcomp}
\usepackage{xcolor}
\usepackage{amsmath}
\usepackage{listings}
\usepackage{algpseudocode}
\usepackage{url}
\usepackage{minted}

\usepackage[ruled,vlined]{algorithm2e}
\DontPrintSemicolon 
\usepackage{makecell} 

\def\BibTeX{{\rm B\kern-.05em{\sc i\kern-.025em b}\kern-.08em
    T\kern-.1667em\lower.7ex\hbox{E}\kern-.125emX}}

\definecolor{darkgreen}{rgb}{0.0, 0.5, 0.0}

\usepackage{pifont}
\begin{document}

\title{Towards An Approach to Identify Divergences in Hardware Designs for HPC Workloads}

\author{
\IEEEauthorblockN{Doru Thom Popovici\IEEEauthorrefmark{1}, Mario Vega\IEEEauthorrefmark{1}, Angelos Ioannou\IEEEauthorrefmark{1}, Fabien Chaix\IEEEauthorrefmark{2}, \\Dania Mosuli\IEEEauthorrefmark{3}, Blair Reasoner\IEEEauthorrefmark{3}, Tan Nguyen\IEEEauthorrefmark{1}, Xiaokun Yang\IEEEauthorrefmark{3}, John Shalf\IEEEauthorrefmark{1}}
\IEEEauthorblockA{\IEEEauthorrefmark{1}Lawrence Berkeley National Lab (LBNL), USA}
\IEEEauthorblockA{\IEEEauthorrefmark{2}Foundation for Research and Technology - Hellas (FORTH), Greece}
\IEEEauthorblockA{\IEEEauthorrefmark{3}University of Houston Clear Lake (UHCL), USA}
}

\maketitle

\begin{abstract}
Developing efficient hardware accelerators for mathematical kernels used in scientific applications and machine learning has traditionally been a labor-intensive task. These accelerators typically require low-level programming in Verilog or other hardware description languages, along with significant manual optimization effort. Recently, to alleviate this challenge, high-level hardware design tools like Chisel and High-Level Synthesis have emerged. However, as with any compiler, some of the generated hardware may be suboptimal compared to expert-crafted designs. Understanding where these inefficiencies arise is crucial, as it provides valuable insights for both users and tool developers. In this paper, we propose a methodology to hierarchically decompose mathematical kernels—such as Fourier transforms, matrix multiplication, and QR factorization—into a set of common building blocks/primitives. Then the primitives are implemented in the different programming environments, and the larger algorithms get assembled. Furthermore, we employ an automatic approach to investigate the achievable frequency and required resources. Performing this experimentation at each level will provide fairer comparisons between designs and offer guidance for both tool developers and hardware designers to adopt better practices.
\end{abstract}


\section{Introduction}
Designing high-performance hardware accelerators remains a labor-intensive endeavor. Substantial engineering effort is required to implement and optimize mathematical kernels such as the Fast Fourier Transform, matrix multiplication, and QR factorization that are widely used in a multitude of scientific applications. Traditionally, most hardware experts rely on low-level hardware description languages (HDLs) like Verilog~\cite{palnitkar2003verilog} or VHDL~\cite{lipsett1989vhdl} to achieve optimal performance and resource utilization for these accelerators. Working at this level not only requires deep domain expertise but also extensive manual tuning to meet timing, performance and area constraints. Moreover, the process of debugging low-level HDL designs and diagnosing performance bottlenecks or functional errors can substantially extend development timelines, thereby exacerbating the challenges associated with delivering efficient accelerators for scientific computing and machine learning applications. As Moore's law is slowing down, there is a growing push towards more productive approaches to deliver efficient hardware designs at lower development costs.

To mitigate these challenges, more user-friendly and higher productivity tools such as high-level synthesis (HLS) frameworks~\cite{VivadoHLS, LegUp, Bambu, scaleHLS} and domain-specific hardware generation tools~\cite{chisel, blueSpec} have gained traction. These solutions aim to increase programming and debugging productivity and reduce time-to-market by allowing developers to describe hardware functionality using high-level abstractions. One prominent example is Chisel~\cite{chisel}, a hardware description language embedded in Scala~\cite{scala_spec} that allows developers to express parameterized hardware designs using Scala’s functional and object-oriented constructs. This approach supports the creation of reusable circuit generators and enables the generation of low-level RTL code through high-level abstractions. This approach allows for significant configurability -- such as variable transform radix sizes, bit precision, core counts, pipeline depths, and unrolling factors—across a broad range of design points. As a result, Chisel makes it easier to explore trade-offs between performance, area, and accuracy. Another widely used approach is Xilinx's Vivado HLS~\cite{VivadoHLS}, which synthesizes RTL code directly from C/C++ descriptions. Developers annotate loops, functions, and data with embedded compiler directives (pragmas) to specify pipelining, loop unrolling, array partitioning, and other architectural constraints, guiding the tool toward a hardware-friendly implementation. It has been demonstrated that HLS can significantly improve programming productivity in many scientific and engineering areas, especially in developing accelerators for compute tasks in graph and data analytics~\cite{HLS_ThunderGP, HLS_GAHLS, HLS_dataAnalytics,HLS_MapReduce}, genome sequencing~\cite{HLS_denovoAsembly, HLS_BQSR, HLS_longreadSeq, HLS_INDEL_realignment, hpcAccel_cpe}, computer vision~\cite{HLS_faceDetection, HLS_FIBLib} and ML/AI inference engines~\cite{HLS_FINN, HLS_systolicCNN, HLS_hls4ml, HLS_CNN_SOC}.

While these high-level tools have enabled faster prototyping and development cycles, there still remains a noticeable gap in performance between auto-generated designs and hand-tuned HDL implementations. Recent comparative studies~\cite{hlsSurvey_Meeus,hlsSurvey_Nane, hls_verilog_Kamkin, chiselSurvey} have attempted to benchmark tools like Chisel, Vivado HLS, and others across various applications, often by developing equivalent full-application designs for each platform. However, these works generally focus on end-to-end performance and fail to analyze the performance of sub-components or explore how each tool handles fundamental primitives and algorithmic choices. Furthermore, they fail to evaluate hand-tuned HLS and auto-generated designs against a baseline from an expert hardware designer in an HDL such as Verilog.  Consequently, their evaluations offer limited insight into why certain tools perform better for particular tasks, and which aspects of the design process contribute most to divergence of the different designs.

In this work, we propose a more fine-grained methodology for comparing hardware generation tools. Rather than evaluating tools at the application level alone, we focus on providing an approach to decompose key mathematical kernels—such as FFT, matrix-matrix multiplication (GEMM), and QR factorization—into common building blocks (e.g., butterfly operations, dot products, AXPY, pointwise operations, and permutations). We then analyze how each tool handles the implementation of these primitives, gradually scaling up to a complete algorithm. We perform this bottom up comparison while automatically attempting to achieve higher frequencies for each of the designs. This bottom-up evaluation framework allows us to identify divergence points in code generation, resource usage, and performance, as well as to assess the impact of parameter tuning at each level. Finally, by automating the synthesis part of the designs we can explore possible implementations for the same design case.

\noindent\textbf{Contributions.} This paper makes the following contributions:
\begin{enumerate}
\item We introduce a structured methodology for comparing hardware generation tools by decomposing mathematical kernels into primitive building blocks.
\item We present a bottom-up comparative analysis, evaluating tool performance, design quality, and tuning behavior at each decomposition level.
\item We develop a semi-automated exploration framework to sweep configuration parameters (e.g., unroll factors, precision types, pipeline depths, target clock frequencies, lane parallelization factor, handshake/interface flavors) for each design block, enabling more comprehensive performance comparisons.
\item We provide an in-depth analysis for the primitives and the GEMM, FFT and QR algorithms. 
\end{enumerate}

\section{Background}
In this section, we briefly describe the three different programming environments by focusing on a simple example. We then briefly talk about the mathematical kernels that were chosen for this work.

\subsection{Tools for Hardware Design}

\begin{figure}[t]
    \centering
    \includegraphics[width=0.3\textwidth]{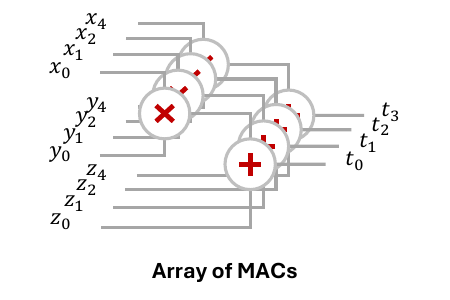}
    \caption{Array of multiply and accumulate operations.}
    \label{fig:mac}
    \vspace{-3mm}
\end{figure}

\begin{figure*}[t]
    \centering
    \includegraphics[width=0.95\textwidth]{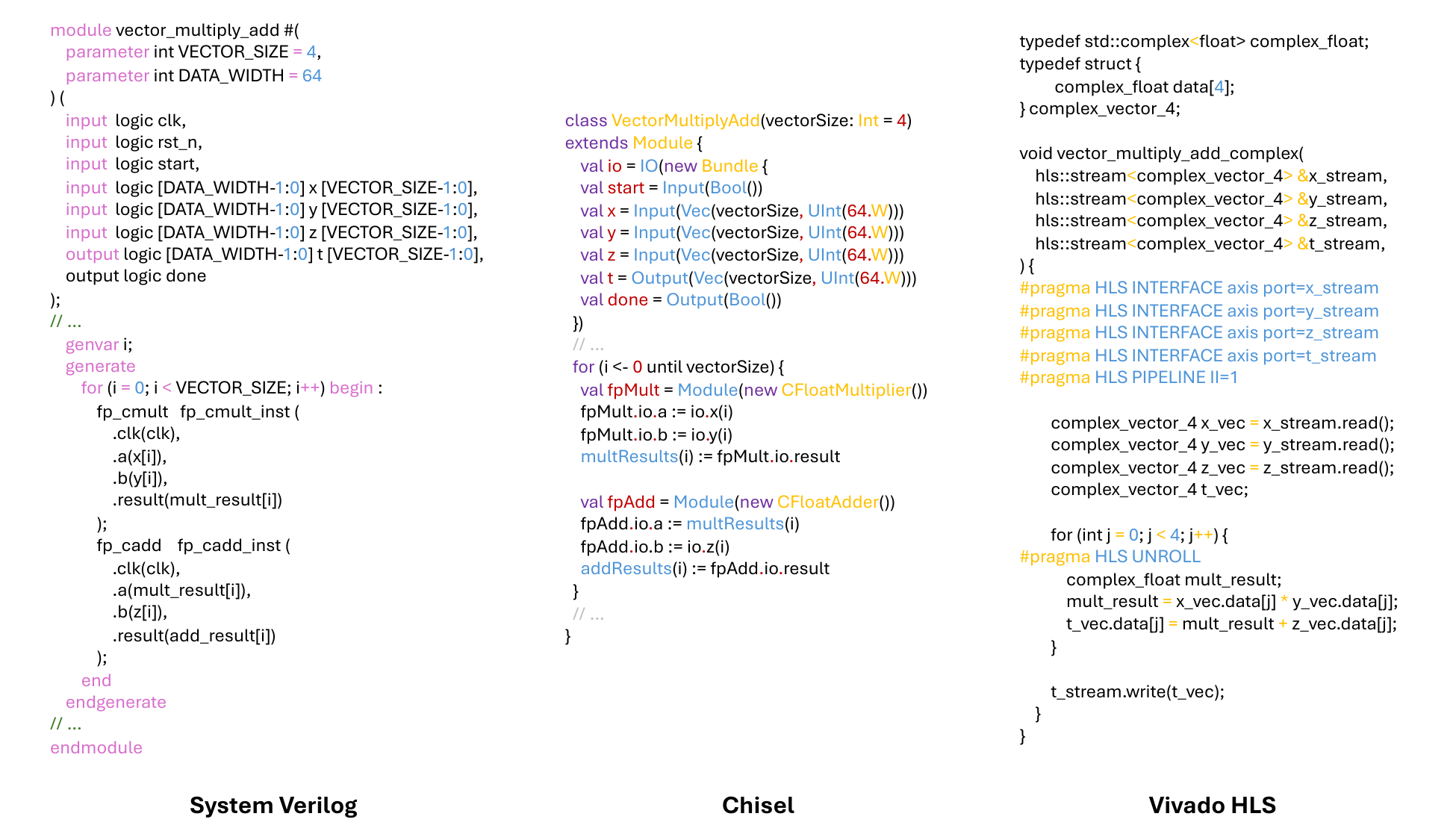}
    \caption{Implementing an array of Multiply and Accumulate (MAC) operations between three vectors $x$, $y$ and $z$ of size $4$ using System Verilog (a), Chisel (b) and Vivado HLS (c). Each element in $x$ is multiplied with the corresponding element in $y$, the result is added to $z$ to produce the final result.}
    \label{fig:tools}
    \vspace{-3mm}
\end{figure*}

We will use a simple mathematical computational pattern as a driving example, namely an array of multiply and accumulate operations as depicted in Figure~\ref{fig:mac}. Given three input arrays $x$, $y$ and $z$ of size $n$, we can define the following operation:
\begin{align}
    t = x\odot y + z,
\end{align}
where $\odot$ is a point-wise multiplication operator that multiplies each element of the $x$ vector with the corresponding elements in the $y$ vector. This is an important computational pattern. For example, if the input vector $z$ is removed, we obtain a basic point-wise calculation defined as
\begin{align}
    t = x\odot y,
\end{align}
which is required by Fourier transforms. Moreover, if the vector $x$ contains duplicates of the element $a$ then the computation reduces to an AXPY operation defined as
\begin{align}
    t = a\cdot y + z,
\end{align}
that is required in linear algebra operations like matrix-vector and matrix-matrix multiplication. The array of multiply and accumulate operations can be easily represented in hardware as depicted in Figure~\ref{fig:tools}. In the following paragraphs, we will briefly  present the implementations of this pattern using the three programming environments, namely Verilog, Chisel and Vivado HLS, outlining their characteristics.

{\bf{Verilog HDL.}} Verilog and its extension SystemVerilog are the de-facto hardware description languages for digital design (For the rest of the paper we will use the term {\em Verilog} to represent both versions). As depicted in Figure~\ref{fig:tools}(a), the array of multiply and accumulates can be written as a Verilog \texttt{module}. The module can be parametrized by specifying the number of units \texttt{VECTOR\_SIZE} and the data width \texttt{DATA\_WIDTH}. Moreover, the module has a bunch of arguments, which represent the inputs and outputs of this hardware unit. Within the module, one must specify the data path and in many cases a state machine that dictates how data flows through the computational units. For brevity, we only present the  module interfaces and instantiations. We use a for loop that will be fully unrolled during elaboration to generate the parallel computation. Within the for loop, a \texttt{fp\_cmult} and a \texttt{fp\_cadd} get instantiated, which correspond to modules that implement  complex multiplication and  complex addition respectively. Each instantiation will require a clock signal and the corresponding inputs and outputs. Note that the output from the multiply unit is passed as input to the addition unit. 

{\bf{Chisel.}} By combining Scala’s functional constructs, object-oriented features, and parameterizations, Chisel offers a productive environment for hardware generator development. Its built-in primitives and utility functions let you express the same design in fewer lines than an equivalent Verilog implementation. In Figure~\ref{fig:tools}(b), we represent the array of multiply-and-accumulates using the Chisel language. The main computation is represented as a class that inherits from the predefined \texttt{Module} class. I/O ports are defined via subclasses of \texttt{Bundle}, which one can customize and instantiate within a module. Similar to the Verilog implementation, a for loop is used to create the \texttt{vectorSize} instances of the multiply and accumulate units. Within the loop, one must instantiate the floating point modules to compute the complex multiplication and complex addition. All the modules must connect between themselves and the input and output wires of the top module. Note that in Chisel, clock and reset signals are not explicitly specified; instead, they are inferred by the compiler when the high-level representation is translated into Verilog.

{\bf{Vivado HLS.}} HLS is a language that uses \texttt{pragma}s to decorate C/C++ function as depicted in Figure~\ref{fig:tools}(c). The array of multiply and accumulates is specified as a C++ function that takes as input/output four arguments of type \texttt{hls::stream}. The type \texttt{hls::stream} is a predefined type by the HLS compiler. A set of \texttt{pragma}s decorate the inputs/outputs. In this example, the input/output arguments are described to use the AXI Stream interface. An additional \texttt{pragma} is used to specify that this unit should be pipelined with an initiation interval (II) of one. In other words, the hardware implementation should expect new data coming every cycle. Finally, the computation is defined using a for loop that is marked for unrolling (replicate the computation). Within the loop the computation is defined as regular C++ multiplication and addition for complex numbers. Note that the HLS implementation is fully specified, no extra information about the state machine is required. The compiler will infer this alongside the clock signal and will create the necessary flow. The \texttt{pragma}s will guide the compiler to the most appropriate decisions.

\subsection{From Primitives to Mathematical Kernels}
The implementations of the previously defined example are quite simple, as one could attest from Figure~\ref{fig:tools}. Things get more complicated when describing full algorithms as the ones defined below.

\begin{figure*}[t]
    \centering
    \includegraphics[width=0.75\textwidth]{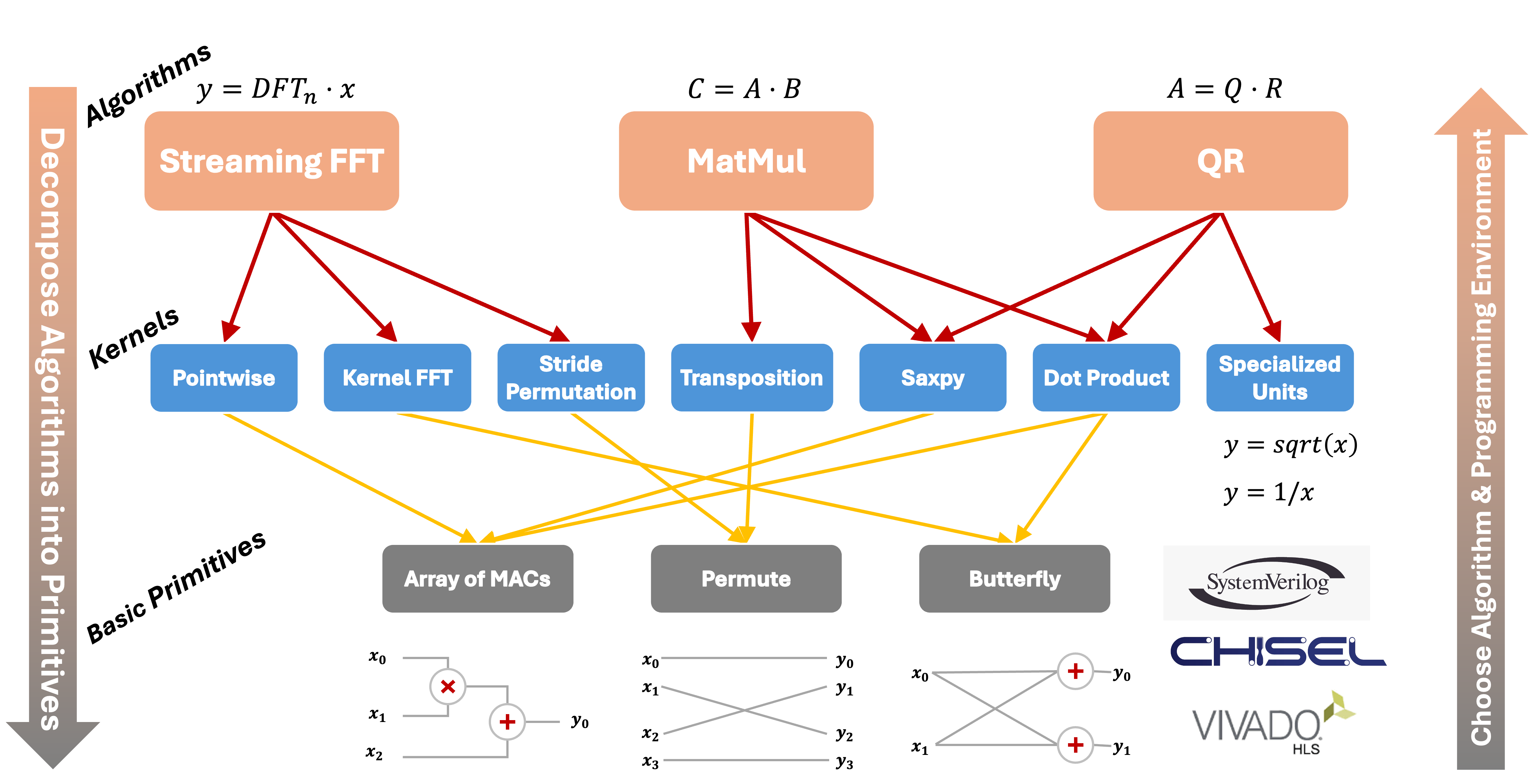}
    \caption{Our overall approach decomposes kernels such as Fourier transforms, matrix multiplication, and QR factorization into primitives, implemented with multiply–accumulate arrays, permutation units, or butterfly units. The designs are realized using three programming tools (Chisel, Verilog, and Vivado HLS) and automatically tested for resource utilization and frequency.}
    \label{fig:mm_primitives}
    \vspace{-5mm}
\end{figure*}

{\bf{Matrix-Matrix Multiplication}} is one of the most widely used mathematical kernels. It appears in a wide range of applications from scientific computing and machine learning. The matrix-matrix multiplication is defined as
\begin{align}
    C = A\cdot B,
\end{align}
where $C\in\mathcal{C}^{m\times n}$, $A\in\mathcal{C}^{m\times k}$ and $B\in\mathcal{C}^{k\times n}$. We use the complex data type for this operation. There are multiple ways to implement the matrix multiplication, each with their own benefits, i.e. dot-product approach, row-scaling approach and outer-product approach.

{\bf{Discrete Fourier Transform}} is used in machine learning, signal processing and scientific simulations. It is defined as a matrix-vector multiplication defined as
\begin{align}
    y = DFT_n\cdot x,
\end{align}
where $DFT_n\in\mathcal{C}^{n\times n}$ is the Fourier complex matrix and $x$ and $y$ are complex column vectors of size $n$. The Fourier computation is never done as a matrix-vector multiplication. Algorithms like Cooley-Tukey~\cite{CooleyTukey}, The Pease variant of the FFT algorithm~\cite{pease1968adaptation} is used to decompose the computation and reduce the $O(N^2)$ to $O(N\log(N))$.

{\bf{QR Factorization}} is a method used in a multitude of solvers to decompose a rectangular matrix into two matrices such that
\begin{align}
    A = Q\cdot R,
\end{align}
where $A\in\mathcal{R}^{m\times n}$, $Q\in\mathcal{R}^{m\times n}$ and $R\in\mathcal{R}^{n\times n}$. The $R$ matrix is upper triangular matrix. Implementing this algorithm is more involved, and Householder projections can be employed to ensure numerical stability. For simplicity, we restrict our focus to real data points rather than complex values. In the following section, we describe our approach to decomposing the computation, selecting an algorithm, implementing the designs in three programming languages, and automatically testing them.

\section{Top-Down Decomposition and Bottom-Up Comparisons}
In this section, we focus on our approach of decomposing the algorithms and automatically testing the different designs at multiple levels. The overview of our approach can be seen in Figure~\ref{fig:mm_primitives}. We will talk about the three algorithms, outlining the implementations and their basic primitives. Finally, we will talk about the automation part of our approach.

\subsection{Decomposing the Algorithms to Primitive Blocks}

Typically, large applications are not implemented as monolithic blocks. They are decomposed into smaller building blocks that are linked together and looped around. The decomposition may differ from one algorithm to another. In the following, we will provide the chose implementations for our benchmarking experiments. We will start with the matrix multiplication, continue with the Fourier transform and conclude with the QR factorization.

{\bf{Matrix-Matrix Multiplication.}} There are multiple implementations for a matrix-matrix multiplication, e.g. dot-product, outer-product, row scaling. For this work we focus on the implementation that scales the columns of the $A$ matrix by elements of the $B$ matrix. Let us assume that we have two input matrices $A\in\mathcal{C}^{2\times 2}$ and $B\in\mathcal{C}^{2\times 16}$. Multiplying the two matrices produces a resultant matrix $C\in\mathcal{C}^{2\times 16}$ such as
\begin{align}
    &\textbf{for } i \leftarrow 0 \textbf{ to } 2 \nonumber \\
    &\quad \textbf{for } j \leftarrow 0 \textbf{ to } 16 \nonumber \\
    &\qquad C[i, j] = 0 \nonumber\\
    &\qquad \textbf{for } k \leftarrow 0 \textbf{ to } 2 \nonumber \\
    &\qquad\quad C[i, j] += A[i, k] \cdot B[k, j]
\end{align}
We improve computation by performing a loop interchange between the $i$ and $j$ loops, and then unrolling the $i$- and $k$-loop. Finally, we reorganize the instructions such that
\begin{align}
    &\textbf{for } j \leftarrow 0 \textbf{ to } 16 \nonumber \\
    &\quad t_{00} = 0 + A[0, 0] * B[0, j]\nonumber\\
    &\quad t_{01} = 0 + A[0, 1] * B[1, j]\nonumber\\
    &\quad t_{10} = 0 + A[1, 0] * B[0, j]\nonumber\\
    &\quad t_{11} = 0 + A[1, 1] * B[1, j]\nonumber\\
    &\quad C[0, j] = t_{00} + t_{01}\nonumber\\
    &\quad C[1, j] = t_{10} + t_{11}
\end{align}

\begin{figure}[t]
    \centering
    \includegraphics[width=0.35\textwidth]{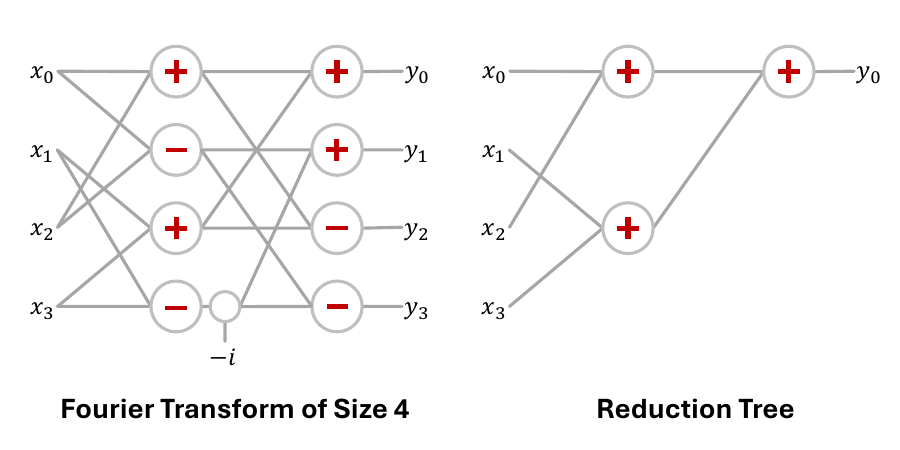}
    \caption{This is the family of butterfly units. The left figure outlines a Fourier transform of size $4$, where all the computation is represented by additions and subtractions. The middle figure replaces all the subtractions with additions, to obtain an all-reduce algorithm. Finally the right figure represents a reduction tree typically used in dot-products.}
    \label{fig:butterfly}
    \vspace{-3mm}
\end{figure}

\begin{figure*}[t]
    \centering
    \includegraphics[width=\textwidth]{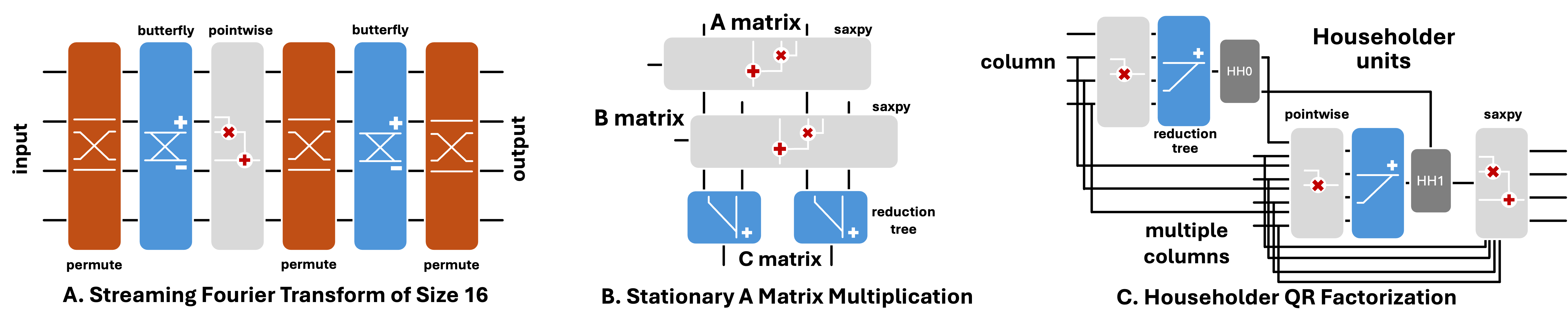}
    \caption{The structure of the three main algorithms under evaluation: 1) the streaming Fourier transform, 2) the matrix-matrix multiplication and 3) the QR factorization. The three figures outline the decomposition of each algorithm using the defined primitives.}
    \label{fig:algorithms}
\end{figure*}

Note that the elements of the $A$ matrix are loaded once and they are reused for all elements of $B$. Moreover, the elements of $B$ are broadcast across multiple instructions (elements of $B$ are reused). As such, we can group the operations to create the so-called AXPY operations, where $B[0, j]$ and $B[1, j]$ are used as scaling constants applied on different columns of the $A$ matrix. Finally a reduction step is performed at the end to compute the final $C$ matrix results. This in turn gives us the following implementation
\begin{align}
    &\textbf{for } j \leftarrow 0 \textbf{ to } 16 \nonumber \\
    &\quad t_{0}[:] = AXPY(B[0, j], A[:, 0], 0)\nonumber\\
    &\quad t_{1}[:] = AXPY(B[1, j], A[:, 1], 0)\nonumber\\
    &\quad C[:, j] = reduce(t_{0}[:], t_{1}[:]),
\end{align}
where $(:)$ represents Matlab notation to specify that the operation is applied for all the elements in a given dimension. The reduction operation is implemented using a reduction tree as defined in Figure~\ref{fig:butterfly} (right). Finally, the described implementation of the matrix multiplication can be visualized in Figure~\ref{fig:algorithms}(b). We use this implementation and its primitives with varying dimensions to benchmark the three different programming languages.

{\bf{Discrete Fourier Transform (DFT).}} There are a myriad of implementations for the Fourier transform. For this work, we focus on the streaming Fourier transform~\cite{} which works on streams of data. Typically a large Fourier transform of size $n$ applied on an input vector $x$ to produce an output vector $y$ is decomposed into smaller transforms given $n = n_0 \times n_1$ and some extra operations. For example, a Fourier transform of size $n = 16$ can be decomposed for $n=4\times 4$ as
\begin{align}\label{eq:ct}
    \tilde{y} = (DFT_4\cdot\left(Twid_{4\times 4}\odot\left(DFT_4\cdot\tilde{x}^T\right)\right)^T)^T,
\end{align}
where $\tilde{x}$ and $\tilde{y}$ are the two-dimensional matrix representations of size $4\times 4$ of the one dimensional input $x$ and output $y$, respectively. Equation~\ref{eq:ct} clearly describes the steps of the Cooley-Tukey algorithm. The first step applies the $DFT_{4}$ in the columns of the transposed matrix $\tilde{x}^T$. The result of the first stage is then point-wise multiplied ($\odot$) with the twiddle matrix $Twid_{4\times 4}$ (this is a matrix that contains the roots of unity). The result is then transposed and a second DFT of size $4$ is applied in the columns. The result is transposed one more time to obtain the final result, stored into the output buffer.

\begin{figure}[t]
    \centering
    \includegraphics[width=0.35\textwidth]{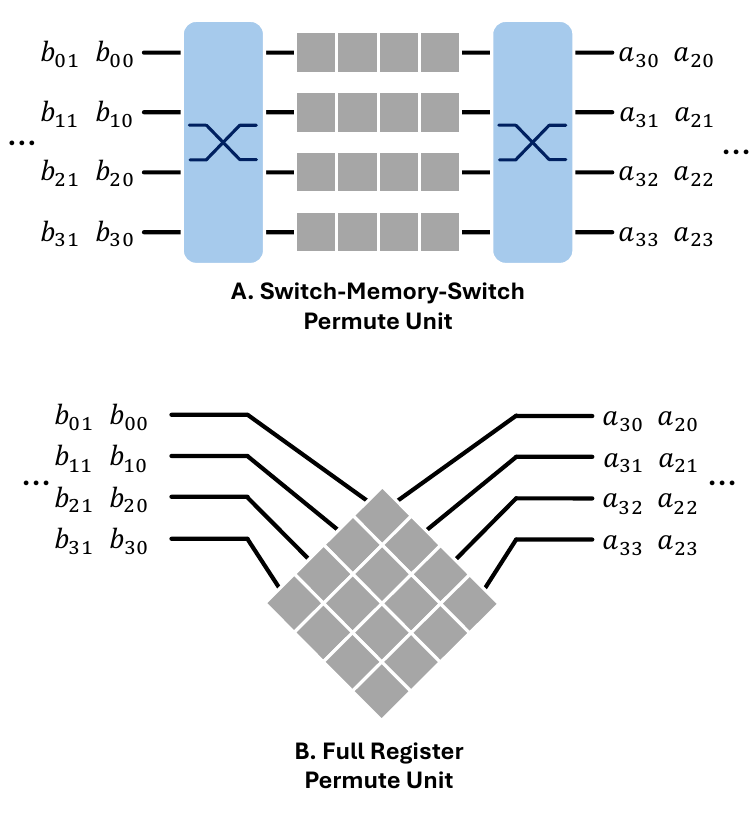}
    \caption{Two implementations for performing a transposition of a matrix of size $4\times 4$, when data is streamed $4$ elements at a time. The top implementation requires a switch-memory-switch~\cite{milder0, milder1}, while the bottom implementation requires $16$ registers.}
    \label{fig:permutation}
    \vspace{-4mm}
\end{figure}

The algorithm described in Eq~\ref{eq:ct} assumes that all the elements of the input are present. However, the algorithm can be manipulated such that the data is streamed into the DFT at the granularity of $4$ elements at a time or per cycle. A streaming DFT of size $16$ requires four cycles to feed the entire input data, and it does not have any feedback loops. Pictorially, the streaming DFT of size $16$ is outlined in Figure~\ref{fig:algorithms}(a). The implementation requires six blocks, namely three permutation units, two butterfly units, and one point-wise multiplication. The permutation units transpose the data. Given that the data comes in streams, internally it has to buffer the data before reshaping it. The permutation is implemented using the details presented in Figure~\ref{fig:permutation}, either using a switch-memory-switch permutation~\cite{milder0, milder1} or using a full register implementation~\cite{jarvinen}. The butterfly unit is just a fully unrolled Fourier transform of size four in this case. The butterfly units expect four elements per cycle and produce four elements per cycle. A pictorial representation of a Fourier transform of size $4$ is outlined in Figure~\ref{fig:butterfly}(a). Finally, the point-wise multiplication is just four complex multiplications to perform the twiddle scaling. For more details about streaming Fourier transforms and Fourier transforms in general, we recommend the reader peruse the following works~\cite{vega2024hpec, spiral_fft, spiral_milder}. We will use the streaming Fourier implementations for large sizes and the corresponding building blocks to test the three different programming languages. 

{\bf{QR Factorization.}} Similar to all other algorithms, the QR has multiple implementations. The QR implementation that uses Householder projectors~\cite{Rafique2012TSQR} has been shown to be the most numerically stable. We will use this implementation for our experiments. Algo~\ref{algo:tsqr} presents the pseudo-code for Householder implementation of the QR factorization for an input matrix $A\in\mathcal{R}^{8\times 4}$. In the proposed algorithm, we do not store the $Q$ matrix, we just keep the upper triangular matrix $R$, which will overwrite the values of the original input $A$.
\begin{algorithm}[h]\label{algo:tsqr}
\caption{Pseudocode for Householder QR}
\For{$k = 0$ \KwTo $3$}{
  $x_k = A[k:7,\,k]$ \\  
  $d_1 = x'_k \cdot x_k$ \\
  $d_2 = \sqrt{d_1}$ \\
  $v_k = x_k$ \\
  $v_k[0] = x_k[0] + \operatorname{sign}(x_k[0])\, d_2$ \\
  $d_3 = d_1 - x_k[0] * x_k[0] + v_k[0] * v_k[0]$ \\
  $\tau_k = -2/d_3$ \\
  \For{$j = k$ \KwTo $3$}{
    $y_j = A[k:7,\,j]$ \\
    $d_4 = v'_k\cdot y_j$ \\
    $d_5 = \tau_k \cdot d_4$ \\
    $y_j = d_5 * v_k + y_j$ \\
    $A[k:7,\,j] = y_j$ \\
  }
}
\vspace{-0.2cm}
\end{algorithm}

The algorithm consists of two \texttt{for} loops. The outer loop iterates through each of the matrix columns. For each column, we first apply the Householder transformations to obtain the corresponding Householder projector. This step consists of a dot-product, square root operation, a sign manipulation and reciprocal. The dot product can be obtained by performing a point-wise multiplication followed by a reduction tree. The inner loop, uses the computed Householder projector to update the other columns of the $A$ matrix and zero out all the elements that are below the the main diagonal. The Householder array is projected onto each column in $A$ by a dot-product. The result scales the projector and accumulates it using an AXPY operation updating each column in $A$. The process is repeated until all columns are fully processed. Note that, in each iteration, the column size decreases by one. Pictorially, the computation is depicted in Figure~\ref{fig:algorithms}(c).
The implementation consists of two point-wise blocks, two reduction-tree blocks, one AXPY block, and two units that perform square root, reciprocal, and additional multiplication operations.
For all our experiments, we implement the Householder QR factorization, for various matrix sizes. Typically we focus on rectangular matrices where the number of rows is twice the size of the number of columns. Moreover, we make the assumption that implementation consumes the entire column, and for now we do not implement an approach that operates on smaller number of elements at a time.

\subsection{Testing Harness and Automated Approach}

As specified in the previous subsection, we choose three algorithms that are widely used in scientific applications as benchmarks for three different hardware programming languages. For each algorithm, we have specified the key building blocks that are needed. As such, we first create benchmarks that focus on the primitives. Table~\ref{tab:benchmarks} outlines the key characteristics of each primitive. We then use the primitives to assemble the larger algorithms as depicted in Figure~\ref{fig:algorithms}. The lower part of Table~\ref{tab:benchmarks} summarizes the key characteristics for each algorithm. For now, we have implemented the primitives and the algorithms by hand. We have focused on efficient Verilog implementations, we provided a similar Chisel implementation and finally we have made use of the features offered by HLS for all primitives and algorithms. We have developed an automated framework that enables seamless integration of different designs, their synthesis on an FPGA, and the collection of key metrics related to frequency and resource utilization. This framework can be run on a supercomputer or server cluster using the provided Slurm script, or on a single server via a shell-based version.

\begin{footnotesize}
\noindent
\begin{table}[t]
\caption{\label{tab:benchmarks}
Table containing the relevant benchmarks, first the primitives and then the actual algorithms. For each benchmark we specify the key characteristics.
}
\begin{tabular}
{p{0.8in}|p{0.45in}|p{0.2in}|p{0.3in}|p{0.85in}}
\hline
Benchmark
& Type & In & Out & Sizes
\\
\hline
Array of MACs
& primitive
& $3$ 
& $1$ 
& 2 -- 16
\\
Fourier Butterfly
& primitive
& 1
& 1
& 2 -- 16
\\
Permutation Unit
& primitive
& 1 
& 1
& 2 -- 16
\\
MatMul
& algorithm
& 2 
& 1
& $2\times 2$ -- $16\times 16$
\\
DFT 
& algorithm
& 1
& 1
& 4, 16, 64, 256
\\
QR Factorization
& algorithm
& 1
& 1
& $16\times 8$ -- $64\times 32$
\\
\hline
\end{tabular}
\vspace{-2ex}
\end{table}
\end{footnotesize}

\begin{figure}[t]
    \centering
    \includegraphics[width=0.35\textwidth]{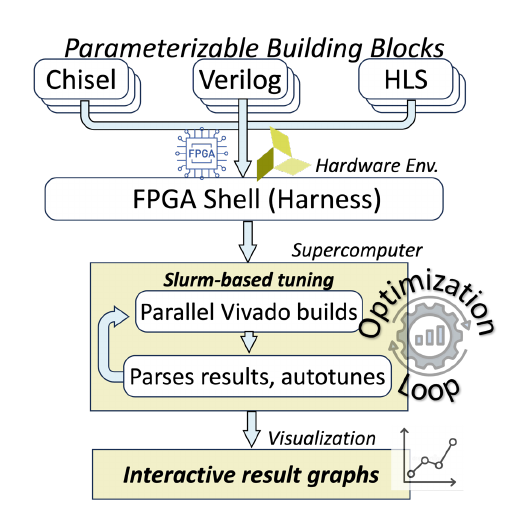}
    \caption{The toolflow for autotuning and comparing hardware accelerator modules launches hardware generation tools on a supercomputer -- iteratively evaluating interim results until the highest achievable performance is reached.}
    \label{fig:tool_overview}
    \vspace{-4mm}
\end{figure}

To evaluate the implementations in a systematic manner, we developed an automation tool depicted in Figure~\ref{fig:tool_overview} that generates the FPGA bitstream and implementation reports using the AMD Vivado toolset. Our tool iteratively tries to find the maximum operating frequency at which a given RTL description and configuration can run without timing violations. In practice, the tool performs the following steps:
\begin{enumerate}
    \item Generate a complete RTL design, wrapping the Unit Under Test (UUT),  for a given operating frequency and configuration (number of lanes, generic parameters)
    \item Execute synthesis, implementation and appropriate reports generation in Vivado, using default Vivado strategies.
    \item Parse reports and select the next target frequency based on measured Worst Negative  Slack (WNS).
    \item Go back to Step 1 until one of the stopping criteria is met.
\end{enumerate}
The next target frequency is based on what we call \textit{achievable frequency} $f_a$. For a given iteration $i$, achievable frequency is the maximum frequency at which the unit is guaranteed to work by the synthesis and implementation tool, and it is typically quite different from the current/target frequency $f$, following Equation~\ref{eq_achievable_freq}: 

\begin{equation}
    f_{a}[i]=\frac{f[i]}{1-\textbf{wns}[i]\cdot{}f[i]}
    \label{eq_achievable_freq}
\end{equation}

The difference between target and achievable frequency is given by $\Delta{}f[i]=\left|f_a[i]-f[i]\right|$. The next target frequency is generated stochastically to deal with uncertainties pertaining to the synthesis and implementation tools:
\begin{equation}
    f[i+1]=\begin{cases}
        & \texttt{if } \textbf{wns} > 0 \texttt{ then } f_a[i]+\mathcal{U}(0,F) \\
        & \texttt{else } f_a[i]+\Delta{}f[i]\cdot\mathcal{U}(-\alpha,\alpha)   
    \end{cases}
    \label{eq_next_freq}
\end{equation}

Iterations stop when one of the following criteria is met:
\begin{enumerate}
\item The maximum number of iterations is reached
\item The maximum frequency supported by the system is reached (i.e. $f_a[i]>= F_{MAX}$).
\item The achievable frequency for the last few iterations is fairly stable, i.e., 
$\frac{min_{j\in[i-W,i]} f_a[j]}{min_{j\in[i-W,i]} f_a[j]} < R $    
\end{enumerate}

In the above equations, constants $\alpha$, $F$ and $R$ have been chosen arbitrarily. 

A set of functionalities has been built around this core to improve usability. It enables users to generate all the necessary files and initiate a Vivado GUI instance so that users may debug and analyze interactively specific instances. In addition, the results of each iteration is appended to a results CSV file. An interactive 2D scatterplot has also been developed to read this file and allows for fast quantitative and qualitative analysis of the results.

\section{Results and Discussion}
In this section, we present our methodology and experimental results. We present the setup of our experiments and outline the key characteristics for each benchmark. We then present the comparison results for each primitive and algorithm using the three different programming languages. We provide in-depth discussion outlining the differences between each design.

\subsection{Methodology}

For all benchmarks, primitive blocks and full algorithms, we used System Verilog, Chisel $6.0.0$ and Vivado HLS $2024.1$. We used the Vivado $2024.1$ tool chain  to synthesize and gather results and targeting an AMD Alveo U280 board. The automation tool was built using Python $3.6$. The tool uses \texttt{json} files to specify the different configurations, specialized \texttt{tcl} files to create the top module that controls all benchmarks and Pandas to gather and visualize the results. The automation tool enabled us to test and extract results for approximately $1500$ implementations for all the mentioned benchmarks. Furthermore, with a push of a button, the tool provides an interactive menu to scroll through the results. In the following subsection, we will present a representative subset of the results. We plan to make the experiments available using a Github repository.

\subsection{Experimental Results and Discussion}

\begin{figure*}[ht]
    \centering
    \begin{minipage}{0.85\textwidth}
        \centering
        \includegraphics[width=0.333\textwidth]{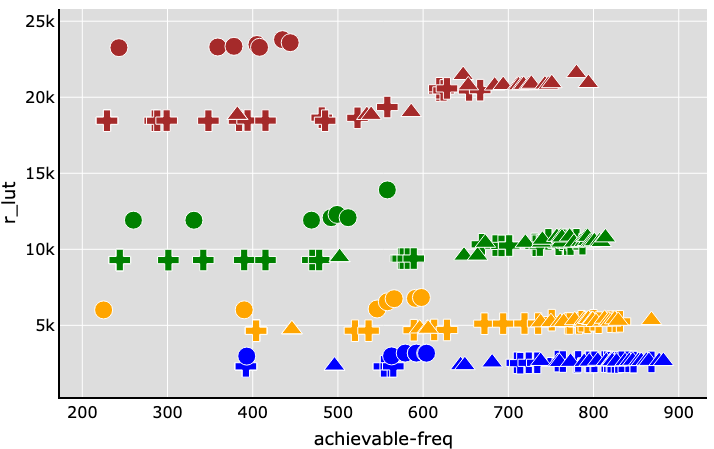}%
        \includegraphics[width=0.333\textwidth]{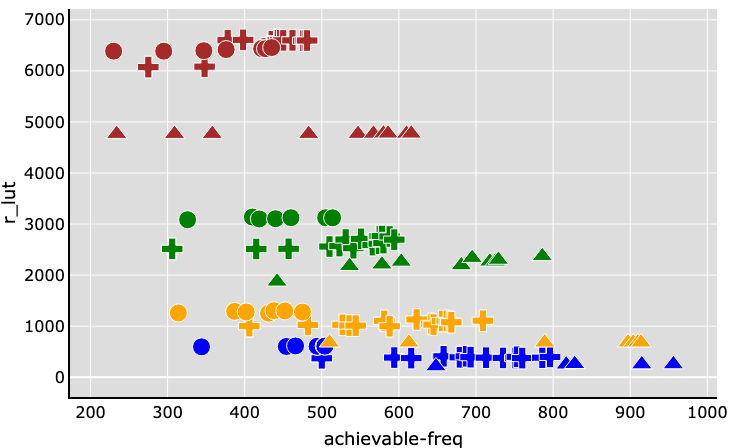}%
        \includegraphics[width=0.333\textwidth]{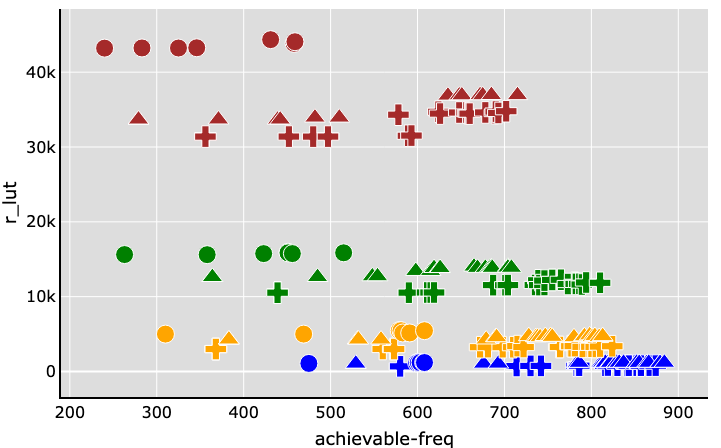}
        
        \vspace{0.02cm}  
        \makebox[0.333\textwidth]{(a) Arrays of MACs}%
        \makebox[0.333\textwidth]{(b) Permutations}%
        \makebox[0.333\textwidth]{(c) Butterflies}
        \vspace{0.15cm}  
        
        \includegraphics[width=0.333\textwidth]{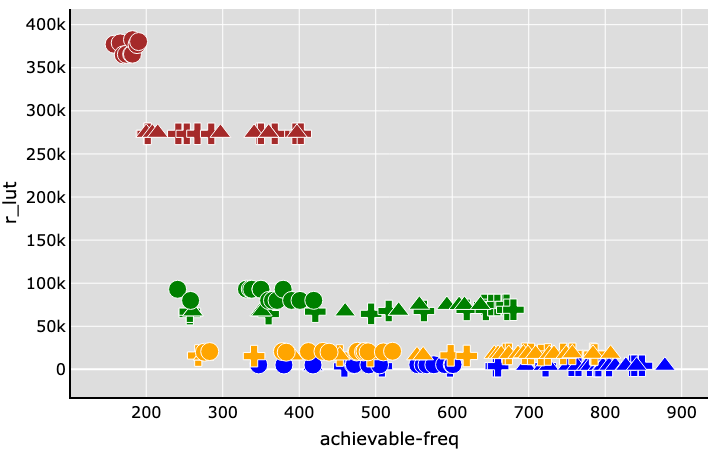}%
        \includegraphics[width=0.333\textwidth]{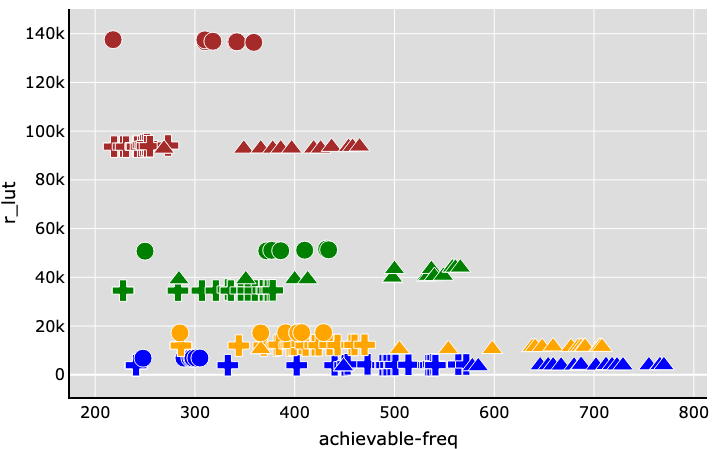}%
        \includegraphics[width=0.333\textwidth]{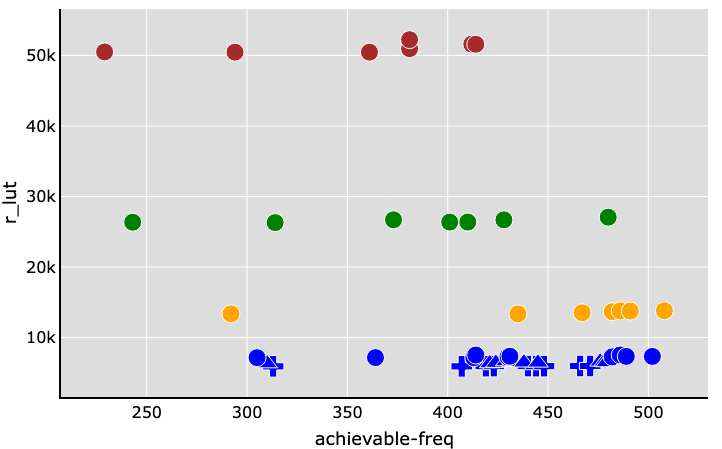}
        
        \vspace{0.02cm}
        \makebox[0.333\textwidth]{(d) GEMMs}%
        \makebox[0.333\textwidth]{(e) FFTs}%
        \makebox[0.333\textwidth]{(f) QR Factorizations}
    \end{minipage}
    \hfill
    \begin{minipage}{0.08\textwidth}
        \centering
        \includegraphics[width=\textwidth]{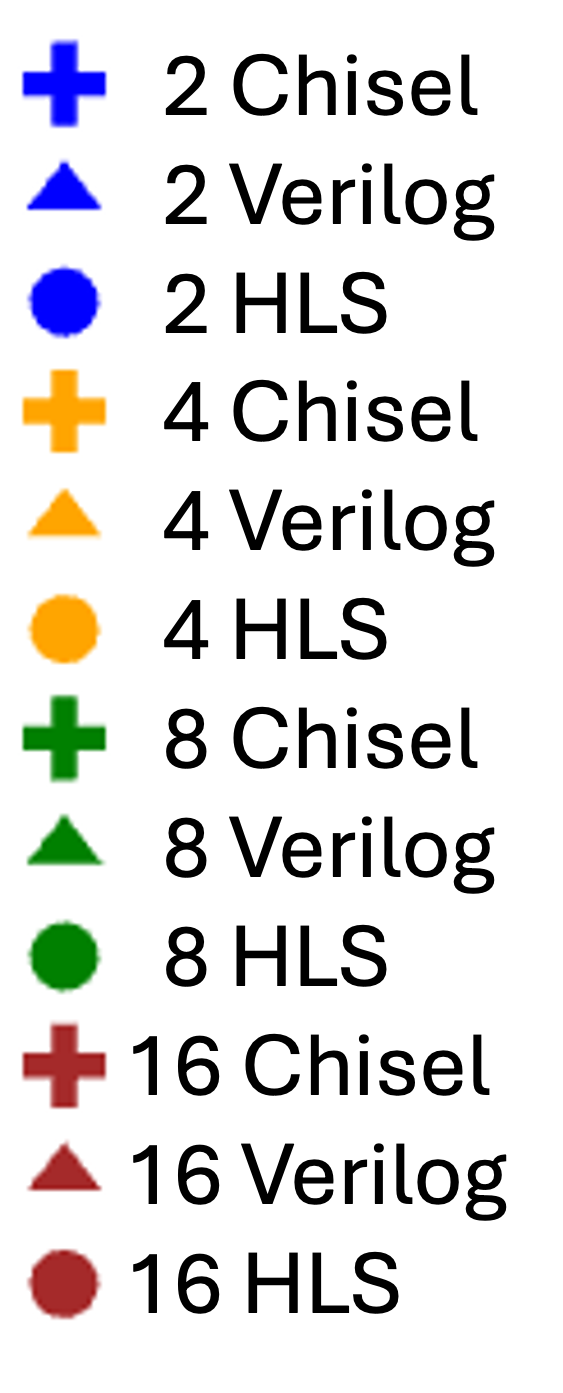}
        \vspace{0.2cm}
    \end{minipage}
    
    \caption{Plots showing the achievable frequency on the $x$-axis and the resource utilization as $LUT$s on the $y$-axis, for our all our benchmarks. The top three plots showcase the results for the three primitives, while the bottom three plots outline the results for the full algorithms. For each plot we provide multiple data points produced by the automation tool for the Verilog, Chisel and HLS implementations. For each experiment, we use different streaming widths ($2$, $4$, $8$, and $16$). Note that for the QR factorization we multiply all streaming widths by a factor of $4$.}
    \label{fig:sevenpanel}
    \vspace{-0.2cm}
\end{figure*}

We begin by presenting our findings and highlighting notable observations from the different experiments. Figure~\ref{fig:sevenpanel} illustrates a subset of the results, with the $x$-axis representing the achievable frequency (in MHz) and the $y$-axis showing resource utilization in terms of LUTs. Due to page limitations, we focus here on these specific metrics; however, the automation tool provides an interactive environment that allows exploration of other parameters such as flip-flop count, BRAM usage, and more. Results are shown for the primitives as well as for the three algorithms. The primary configuration parameter is the streaming width, which defines the total number of lanes per input and output. For the primitives, GEMM, and FFT implementations, we evaluate streaming widths of $2$, $4$, $8$ and $16$ complex numbers—corresponding to $128$, $256$ $512$ and $1024$ bits, respectively. For the QR implementation, each of these four streaming widths is multiplied by $256$ bits, as the design operates on a large number of elements per column.

The top three plots (a–c) present results for the primitive blocks. The MAC and butterfly units are purely computational: data is streamed in, processed through floating-point adders and multipliers, and results are streamed out. For small streaming widths, all three implementations (Verilog, Chisel, and HLS) consume similar resources, with Verilog consistently achieving higher frequencies. As streaming width increases, Chisel and Verilog remain competitive in resource usage, whereas HLS requires progressively more resources. The permutation unit shows a more pronounced separation between the three approaches: Verilog consistently delivers lower resource usage and high frequencies, with both Chisel and HLS consuming more resources as width grows. 

We extend our evaluation from individual primitives to three full algorithms built upon them, allowing us to assess performance in more complex designs, including GEMM, FFT, and QR factorization. For the GEMM implementation (Figure~\ref{fig:sevenpanel}(d)), Verilog and Chisel show similar resource usage and achievable frequency, while HLS consumes more resources and delivers lower frequency. In GEMM, resource requirements grow quadratically with streaming width due to the chosen algorithm: larger widths require proportionally more data points from the A matrix. This may explain why HLS attempts a more aggressive implementation strategy, albeit at a higher resource cost. For the FFT implementation (Figure~\ref{fig:sevenpanel}(e)), trends mirror those seen in GEMM—Verilog and Chisel remain close in resource usage, with Verilog outperforming in frequency, while HLS again lags behind. The FFT combines permutation units, butterflies, and point-wise scaling operations; although all three implementations perform well on the individual primitives, assembling them into a more complex design appears to present greater challenges for Chisel and HLS. Finally, the QR factorization results (Figure~\ref{fig:sevenpanel}(f)) show similar behavior. Due to delays in generating the larger Verilog and Chisel designs, we only report results for the smallest configuration. Interestingly, all three designs exhibit similar resource usage and achievable frequency. In both the Verilog and Chisel versions, the data path is optimized to use a single dot product unit, exploiting inter-column dependencies to enable reuse. The HLS compiler achieves a comparable optimization: guided by pragmas, it detects redundant dot product operations and generates a single implementation reused across computation stages.

While the results demonstrate that our implementations function correctly across the three languages and varying streaming widths, the primary contribution lies in the methodology itself. Our approach systematically decomposes complex algorithms into reusable primitives, enabling a clear understanding of data flow and computational dependencies. Coupled with the automation tool, this methodology allows users to rapidly explore multiple implementations and configurations, comparing resource utilization and achievable frequency with minimal manual effort. In other words, the main value is not in which of these specific kernels performs better in Verilog, Chisel, or HLS, but in providing a framework where algorithmic decomposition and automated evaluation empower designers to make informed trade-offs. The results shown serve mainly as a proof of concept, illustrating that the process works correctly and can guide decision-making across a wide range of hardware implementations. Furthermore, our methodology enables hardware developers to incorporate new design tools and benchmark them against established implementations.

\section{Conclusion}

\vspace{-0.1cm}
We present a hierarchical framework for algorithm comparison, demonstrated on broadly used HPC e kernels such as Fourier transforms, matrix multiplication, and QR factorization. By decomposing each algorithm into primitives, the framework exposes data flow and computational dependencies, enabling multi-level comparisons across implementations. An accompanying automated tool integrates diverse primitives and implementations, allowing efficient evaluation of nearly 1,500 designs. The framework is readily extensible to other hardware design tools, supporting researchers and developers in exploring alternative algorithmic strategies for future accelerators.

\bibliographystyle{unsrt}
\bibliography{article}

\end{document}